\documentclass[sigconf]{acmart}
\usepackage[most]{tcolorbox}
\usepackage{seqsplit}
\newcommand{\code}[1]{\texttt{\seqsplit{#1}}}
\usepackage{balance}

\usepackage{adjustbox}
\usepackage{array,booktabs}
\usepackage{threeparttable,booktabs,array}
\usepackage{makecell}
\usepackage{multirow}
\newcolumntype{L}[1]{>{\raggedright\arraybackslash}p{#1}}
\AtBeginDocument{%
  }

\copyrightyear{2026}
\acmYear{2026}
\setcopyright{cc}
\setcctype{by}
\acmConference[FSE Companion '26]{34th ACM Joint European Software Engineering Conference and Symposium on the Foundations of Software Engineering}{July 05--09, 2026}{Montreal, QC, Canada}
\acmBooktitle{34th ACM Joint European Software Engineering Conference and Symposium on the Foundations of Software Engineering (FSE Companion '26), July 05--09, 2026, Montreal, QC, Canada}
\acmDOI{10.1145/3803437.3805265}
\acmISBN{979-8-4007-2636-1/2026/07}

\begin{document}

\title{Human-aligned AI Model Cards with Weighted Hierarchy Architecture}

\author{Pengyue Yang}
\affiliation{%
  \department{School of Electrical and Computer Engineering}
  \institution{The University of Sydney}
  \city{Sydney}
  \country{Australia}
}
\email{pyan8493@uni.sydney.edu.au}

\author{Haolin Jin}
\affiliation{%
  \department{School of Electrical and Computer Engineering}
  \institution{The University of Sydney}
  \city{Sydney}
  \country{Australia}}
\email{hjin3177@uni.sydney.edu.au}

\author{Qingwen Zeng}
\affiliation{%
  \department{School of Electrical and Computer Engineering}
  \institution{The University of Sydney}
  \city{Sydney}
  \country{Australia}}
\email{qzen5227@uni.sydney.edu.au}

\author{Jiawen Wen}
\affiliation{%
  \department{School of Electrical and Computer Engineering}
  \institution{The University of Sydney}
  \city{Sydney}
  \country{Australia}}
\email{jiawen.wen@sydney.edu.au}

\author{Harry Rao}
\affiliation{%
  \institution{Bytedance}
  \city{Sydney}
  \country{Australia}}
\email{harry.rao@bytedance.com}

\author{Huaming Chen}
\affiliation{%
  \department{School of Electrical and Computer Engineering}
  \institution{The University of Sydney}
  \city{Sydney}
  \country{Australia}}
\email{huaming.chen@sydney.edu.au}

\renewcommand{\shortauthors}{Pengyue et al.}

\begin{abstract}
The proliferation of Large Language Models (LLMs) has led to a burgeoning ecosystem of specialized, domain-specific models. While this rapid growth accelerates innovation, it has simultaneously created significant challenges in effective model discovery and adoption. Users often struggle to navigate this fragmented landscape due to inconsistent, incomplete, and imbalanced documentation across platforms. 
Existing documentation frameworks, such as Model Cards and FactSheets, have advanced efforts toward standardize reporting, which are still often static, largely qualitative, and not always well suited for rigorous cross-model comparison lacking quantitative mechanisms. This gap exacerbates model underutilization and impedes responsible adoption. To address these gaps, we introduce the Comprehensive Responsible AI Model Card Framework (CRAI-MCF), a novel framework that transitions from static disclosures to actionable, human-aligned documentation. Grounded in Value Sensitive Design, CRAI-MCF is built upon an empirical analysis of 240 open-source projects, distilling 217 parameters into an eight-module, value-aligned architecture. Our framework introduces a quantitative sufficiency criterion to operationalize documentation sufficiency and support more rigorous cross-model documentation comparison within a unified scheme. By integrating technical, ethical, and operational dimensions, CRAI-MCF provides a structured and extensible template for authoring, organizing, and maintaining LLM documentation, with scoring used as a lightweight aid for prioritization and sufficiency checking.
\end{abstract}




\settopmatter{printacmref=false}

\maketitle

\section{Introduction}
\label{sec:intro}

The rapid advancement of Large Language Models (LLMs) has revolutionized numerous domains, addressing complex challenges from natural language understanding to code generation and creative content creation \cite{zhao2023survey}. As of 2025, the proliferation of LLMs has accelerated dramatically, with over 2 million public models hosted on platforms like Hugging Face, reflecting an exponential growth where training compute doubles every five months \cite{laufer2025anatomy, maslej2025artificial}. This high-speed development phase has led to the emergence of various domain-specific LLMs tailored to specialized needs, such as medical models (e.g., BioMedGPT \cite{luo2023biomedgpt} for clinical diagnostics), financial models (e.g. FinGPT \cite{yang2023fingpt} for risk assessment and fraud detection), and time-series models (e.g., Chronos \cite{ansari2024chronos} for forecasting in supply chain and energy sectors). These innovations enable unprecedented efficiency in solving real-world problems, yet they also introduce challenges in model discovery and adoption within the burgeoning LLM ecosystem.

The sheer volume of emerging LLMs creates significant barriers for users, including engineers, researchers, and domain experts, who often lack the time to thoroughly evaluate each model's features, details, and suitability for their specific needs \cite{zhao2023survey}. In practice, users may rely on superficial or potentially misleading evaluation signals, leading to the premature dismissal of potentially valuable models \cite{banerjee2024vulnerability}. The outcome is early exclusion of promising domain-specific models, such as clinical systems, before adequate evaluation occurs \cite{zhang2025revolutionizing}. Documentation gaps may also discourage consideration of models intended for privacy-sensitive settings, where risks and safeguards are harder to assess \cite{shanmugarasa2025sok}. In industrial settings, this inefficiency not only stifles innovation but also amplifies overhead, as teams waste resources rediscovering or rebuilding capabilities that existing models could provide, underscoring a critical conflict between rapid model proliferation and effective user adoption.

Existing mechanisms aim to facilitate understanding of LLMs, including their usage and functional scope, but they fall short in addressing these challenges. For instance, README files offer basic overviews, while Model Cards provide structured disclosures on intended uses, performance metrics, and limitations \cite{mitchell2019model}. Similarly, FactSheets and System Cards extend this to ethical risks and system-level behaviours \cite{arnold2019factsheets,hurst2024gpt,weidinger2021ethical}. Despite these efforts, persistent issues hinder their effectiveness: documentation is often inconsistent across repositories, with key fields missing (e.g., update history, risk mitigations, or fairness metrics); coverage is imbalanced, prioritizing technical details over ethical or operational aspects; and comparability remains limited, lacking quantitative tools for cross-model evaluation. These shortcomings exacerbate fragmentation, making it difficult for users to quickly assess applicability and leading to prolonged audits or suboptimal model selections in practice.

To overcome these inconsistencies, imbalances, and limitations in existing mechanisms, we introduce the Comprehensive Responsible AI Model Card Framework (CRAI-MCF), a novel approach that transitions from static disclosures to actionable, human-aligned documentation practices. Grounded in Value Sensitive Design (VSD), CRAI-MCF consolidates 217 parameters from an empirical analysis of 240 open-source projects into eight value-aligned modules, ensuring balanced coverage across technical, ethical, and operational dimensions while addressing missing fields. It incorporates a hierarchical architecture to enhance navigability and reduce fragmentation, and a weighted scoring system for stakeholder-specific prioritization to mitigate imbalances. This framework supports adaptability in evolving industrial contexts, enabling practitioners to structure, maintain, and review LLM documentation more systematically while upholding ethical and operational integrity. In summary, this work delivers the following research and practical contributions:
\begin{enumerate}
    \item \textbf{Practice-Grounded Parameter Taxonomy}: A comprehensive inventory of 217 semantically atomic parameters, mapped to 14 responsible AI principles, distilled from an empirical analysis of 240 open-source projects across Hugging Face, GitHub, and Kaggle. This taxonomy provides a reusable foundation for consistent documentation in industrial settings, addressing fragmentation and enabling standardized reporting.
    \item \textbf{Value-Aligned Modular Architecture}: An eight-module framework that balances technical, ethical, and operational concerns, reducing cognitive overload and enhancing navigability for engineering teams in real-world workflows.
    \item \textbf{Quantitative sufficiency criterion}: We formalize frequen-cy-based priors and module baselines to operationalize "how much is sufficient" and enable rigorous cross-model comparison under a unified scheme.
    \item \textbf{Practical Artifacts}: A parameter template, a lightweight sufficiency scheme, and a demo prototype that support structured documentation authoring, review, and replication.
\end{enumerate}

\section{Related Work}

The proliferation of Large Language Models (LLMs) has created a burgeoning ecosystem of specialized tools. Domain-specific models are rapidly emerging, such as BioMedGPT for clinical diagnostics, FinGPT for financial risk assessment, and Chronos for time-series forecasting. However, this rapid growth has outpaced the development of consistent and thorough documentation practices. Empirical analyses of this ecosystem reveal significant shortcomings in how these models are documented. For instance, several studies have examined documentation in open-source platforms. Dodge et al. \cite{dodge2019show} found inconsistencies in NLP model cards, with many omitting key details such as training data and evaluation protocols. Bender and Friedman noted similar deficiencies in linguistic resources, where the lack of standardized fields hindered reuse \cite{bender2018data}. More recent audits of Hugging Face repositories confirm this is a persistent problem: while model cards are encouraged, many projects leave critical sections blank, particularly those related to environmental impact, limitations, and evaluation \cite{liang2024systematic}. This evidence highlights a critical gap between the proliferation of models and the clarity required for their responsible adoption.

While these studies diagnose the failures in documentation practice, the problem is compounded by the inherent limitations of existing AI documentation frameworks themselves. Pioneering frameworks sought to establish standards, but they have proven insufficient in practice. For example, while Datasheets for Datasets established a crucial precedent for data transparency, its scope is limited to datasets. Building on this, Model Cards introduced structured templates for model reporting but, in practice, function as static checklists that suffer from inconsistent adoption and are difficult to maintain across versions \cite{castano2024analyzing, bommasani2023foundation}. Subsequent frameworks like FactSheets and System Cards attempted to broaden the scope to AI services and societal risks, but are often hindered by high authoring costs, are burdensome to update, and remain predominantly qualitative. As Table\ref{tab:framework-comparison} indicates, these pioneering efforts are constrained by a shared paradigm: they are static, often imbalanced in their coverage of technical versus ethical issues, and lack the quantitative mechanisms needed for rigorous cross-model comparison. This leaves a clear and urgent need for an actionable engineering toolkit that supports comparable, stakeholder-aligned evaluation, a gap our framework is designed to fill.

\begin{table*}[t]
\setlength{\tabcolsep}{4pt}
\renewcommand{\arraystretch}{1.12}
\scriptsize
\centering
\begin{threeparttable}
\caption{Comparison of AI documentation artifacts (scope, strengths, and industrial limitations).\tnote{*}}
\begin{tabular}{p{0.15\linewidth} p{0.15\linewidth} p{0.26\linewidth} p{0.26\linewidth} p{0.07\linewidth}}
\toprule
\textbf{Artifact} & \textbf{Scope / Target} & \textbf{Primary Strengths} & \textbf{Key Industrial Limitations} & \textbf{Comparability} \\
\midrule
Datasheets for Datasets\cite{gebru2021datasheets} &
Datasets &
Provenance, collection process, consent \& intended use; improves dataset transparency. &
Focuses on data rather than model behavior; limited versioning/lifecycle support. &
\emph{Low} \\
\addlinespace[2pt]
Model Cards\cite{mitchell2019model} &
Trained models &
Template for intended use, metrics, limitations, fairness notes; widely recognized. &
Static in practice; inconsistent adoption across repositories; limited cross-repo comparison. &
\emph{Medium} \\
\addlinespace[2pt]
FactSheets\cite{arnold2019factsheets} &
AI services / systems &
Lifecycle-oriented disclosure to support enterprise assurance. &
Higher authoring/maintenance cost; heterogeneous adoption; little quantitative scoring. &
\emph{Medium} \\
\addlinespace[2pt]
System Cards\cite{hurst2024gpt} &
Foundation/LLM-scale systems &
Societal/ethical risk surfacing; harm taxonomies; mitigation narratives. &
Predominantly qualitative; costly to maintain; narrower fit for small/domain models. &
\emph{Low} \\
\addlinespace[2pt]
AI Usage Cards\cite{wahle2023ai} &
AI use in workflows &
Standardized usage reporting; emphasizes transparency and accountability in deployment. &
Covers use rather than model internals; adoption nascent; limited integration with scoring. &
\emph{Low} \\
\bottomrule
\end{tabular}
\begin{tablenotes}\footnotesize
\item[*] \textbf{Comparability} is an operational ordinal assessment used in this paper (Low = qualitative/ad hoc; Medium = partial structure; High = standardized quantitative scoring). It reflects the extent to which an artifact natively supports cross-project assessment; it is \emph{not} a value judgment on the artifact’s overall merit.
\label{tab:framework-comparison}
\end{tablenotes}
\end{threeparttable}
\end{table*}

\vspace{0.20 cm}

\section{Methodology}
\label{sec:method}
Engineering audits reveal three persistent pain points in current practice: documentation is \emph{fragmented} across READMEs, model cards, and issue trackers; coverage is \emph{imbalanced} (performance over-reported while risk, lifecycle, and accountability are under-reported); and cross-project assessment lacks \emph{comparability}. These weaknesses undermine not only transparency but also regulatory compliance and cross-team collaboration, creating recurring bottlenecks in industrial AI workflows.

This work is organized around three research questions (RQ1--RQ3) that guide our method and evaluation:
\begin{itemize}
    \item \textbf{RQ1:} To what extent can a practice-derived taxonomy resolve inconsistency and coverage fragmentation in current LLM documentation?
    \item\textbf{RQ2:} Does a value-aligned modular architecture with quantitative scoring reduce cognitive load and enable rigorous cross-model comparison across diverse LLMs?
    \item \textbf{RQ3:} Can the framework be operationalized so that authoring and maintenance remain within industrial limits while preserving documentation quality?
\end{itemize}

To avoid an ad hoc fix, we ground our approach in \emph{Value Sensitive Design (VSD)}\cite{friedman2002value}, using its empirical-conceptual-technical investigations as the backbone and translating them into an engineering pipeline. CRAI-MCF operationalizes VSD through three interlocking pillars: (i) \textbf{Practice-Derived Requirements}, a synthesis of widely used guidance and what high-adoption projects already report, yielding \emph{217} semantically atomic parameters consolidated from \emph{240} projects; (ii) \textbf{Hierarchical Design}, an \emph{eight}-module, containment-only structure chosen for navigability and non-overlap in day-to-day workflows; and (iii) \textbf{Scoring System}, simple parameter priors and module baselines derived from corpus frequency and observed coverage, so teams know “what to fill first” and “how much is enough.” Subsequent subsections expand these pillars and their instantiation in practice.

\subsection{Methodological Foundations: Value Sensitive Design}
\label{sec:vsd}

We ground our approach in Value Sensitive Design (VSD)\cite{friedman2002value} and run iterative empirical--conceptual--technical cycles to move from values to operational practice. The goal is not to propose yet another checklist, but to translate widely shared expectations about responsible AI into constraints that real engineering teams can execute and audit.

\paragraph{VSD investigations and artifacts.}
\textit{Empirical.} We studied high-adoption repositories and shadowed day-to-day workflows to surface where documentation breaks down in practice and what information people actually need to make decisions. The output of this stage is a grounded problem framing and a corpus of concrete materials against which any proposal must hold up.

\textit{Conceptual.} Rather than inventing new categories, we compress established norms and guidance into a small set of design constraints that bound the solution space. At this level, the documentation should be easy to navigate during real tasks, state facts in unambiguous and self-contained units, and anchor claims in evidence that others can verify. These constraints are deliberately minimal so they travel across domains and organizational styles.

\textit{Technical.} With the constraints in hand, we shape them into a deployable specification aimed at everyday use: structures that fit existing tooling and review rhythms, fielding that admits measurement and automation, and simple operating rules that make priorities clear and effort bounded. The emphasis is on applicability, repeatability, and auditability rather than narrative flourish.

\begin{figure*}[t]
    \centering
    \includegraphics[width=1.55\columnwidth]{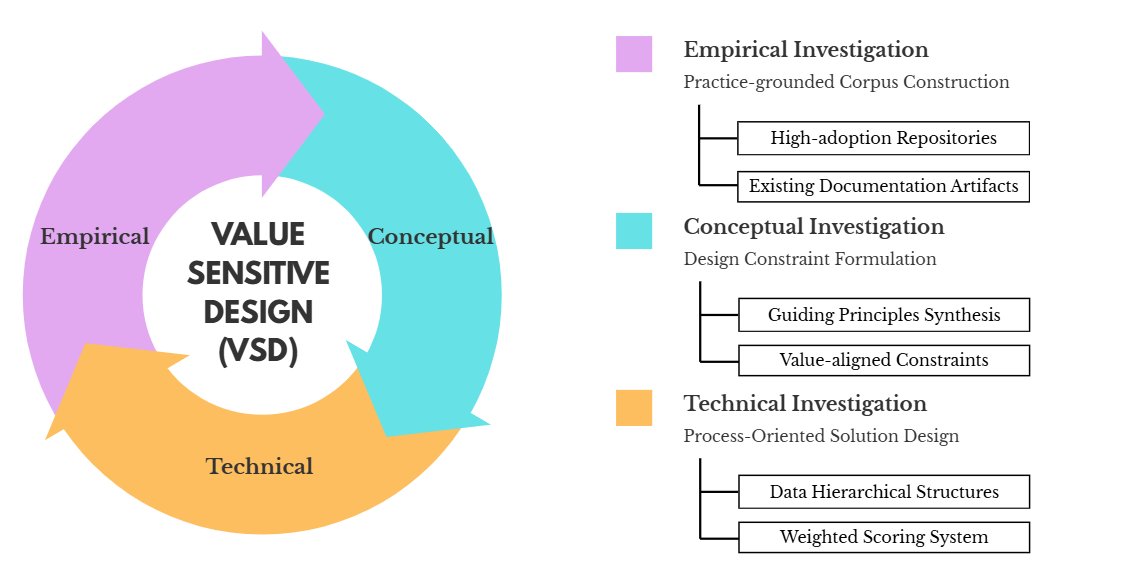}
    \caption{The VSD-anchored research pipeline}
    \label{fig:vsd_process}
\end{figure*}

\paragraph{Design implications.}
Together, these choices reduce context switching during reviews, turn prose into units that can be checked and reused, privilege verifiable artifacts over free text to keep updates lightweight, and keep authoring and maintenance within predictable limits as systems evolve. Figure\ref{fig:vsd_process} sketches the pipeline; subsequent sections instantiate these choices in concrete form.

\subsection{Practice-Derived Requirements}\label{sec:requirements}
\paragraph{Guiding Principles}
To anchor our analysis in established norms for responsible AI, we first synthesized a set of guiding principles from 14 highly influential sources. These included industry standards (e.g., Model Cards\cite{mitchell2019model}, FactSheets\cite{arnold2019factsheets}, Datasheets for Datasets\cite{gebru2021datasheets}), international guidelines (e.g., OECD AI Principles) \cite{yeung2020recommendation}, and prominent risk management frameworks (e.g., NIST AI RMF\cite{ai2023artificial}). 
Instead of merely adopting one standard, we performed a thematic synthesis to consolidate overlapping concepts and identify distinct areas of focus. 

This process resulted in the 14 guiding principles summarized in Table\ref{tab:guiding_principles}. 
These principles collectively span the AI lifecycle-from data provenance and model development to deployment considerations and accountability-and served as a conceptual scaffold for identifying and categorizing documentation parameters in our empirical analysis.

\begin{table}[t]
\setlength{\tabcolsep}{4pt}
\renewcommand{\arraystretch}{1.1}
\scriptsize
\centering
\caption{14 Responsible AI Principles guiding CRAI-MCF parameter selection, grouped into three categories.}
\label{tab:guiding_principles}
\begin{tabular}{p{0.18\linewidth} p{0.55\linewidth} p{0.10\linewidth}}
\toprule
\textbf{Principle} & \textbf{Focus} & \textbf{Source} \\
\midrule
\multicolumn{3}{l}{\textbf{Correctness of Use (1–6)}} \\
1. Scientific rigor & Evidence-based, validated documentation. & \cite{mandel2022policy} \\
2. Uncertainty explanation & Transparent limits, biases, error margins. & \cite{mandel2022policy,mitchell2019model} \\
3. User/usage clarity & Clear target users and scope. & \cite{mandel2022policy,linke2017design} \\
4. Logical reasoning & Documentation grounded in explicit rationale. & \cite{mandel2022policy} \\
5. Ethical considerations & Address fairness, privacy, compliance. & \cite{yeung2020recommendation,ai2023artificial, weidinger2021ethical} \\
6. Unambiguous language & Standardized and precise terminology. & \cite{mandel2022policy,chmielinski2024clear} \\
\midrule
\multicolumn{3}{l}{\textbf{Restrictions / Instructions of Use (7–9)}} \\
7. Source transparency & Cite datasets, provenance, and methods. & \cite{gebru2021datasheets} \\
8. Model consistency & Documentation aligned with model goals. & \cite{arnold2019factsheets} \\
9. Timeliness & Ensure up-to-date documentation and versions. & \cite{winecoff2025improving} \\
\midrule
\multicolumn{3}{l}{\textbf{Portability of Use (10–14)}} \\
10. Developer evaluation & Capture developer insights and risks. & \cite{weisz2024design} \\
11. Alternatives availability & Provide dataset/model alternatives. & \cite{weisz2024design} \\
12. Visualization & Use visuals to aid interpretation. & \cite{weisz2024design} \\
13. Interactivity & Support engagement and feedback loops. & \cite{weisz2024design} \\
14. Scalability & Adapt to evolving needs and contexts. & \cite{winecoff2025improving,batool2023responsible} \\
\bottomrule
\end{tabular}
\end{table}

\paragraph{Project Corpus Selection}
We adopted a purposive, \textbf{rank-based} approach rather than random sampling to capture mature, actively maintained documentation. Between \textbf{May~2023 and July~2025}, we built a candidate pool from platform-native popularity lists—GitHub "trending"/star leaderboards, Hugging Face "most downloaded/ trending" model lists, and Kaggle "most voted/hot" pages. From this pool, we retained entries meeting at least one per-platform adoption signal (GitHub \emph{star} $\geq$ 500 or \emph{fork} $\geq$ 100; Hugging Face \emph{download} $\geq$ 10{,}000; Kaggle \emph{upvote} $\geq$ 50) and verified recency against a uniform snapshot date (\textbf{2025-07-12}): each retained item shows activity within the prior 12 months (a commit, tagged release, or model-card update).

\textit{De-duplication.} We removed forks, mirrors, and template repositories (via GitHub network metadata and README heuristics), collapsed cross-platform duplicates (same canonical model family explicitly cross-referenced across GitHub/Hugging Face), and merged packaging-only variants. When multiple artefacts co-existed, we preferred the repository that contains training/evaluation artefacts and primary documentation over downstream wrappers or mirrors. This yielded a corpus of \textbf{240} unique, high-adoption AI projects.

\paragraph{Parameter Synthesis and Normalization}
A core challenge in creating a comprehensive inventory from diverse sources is to establish a consistent, fundamental unit of analysis. We therefore define a \textbf{parameter} as \textbf{any discrete, self-contained unit of documentation that addresses a specific question about an LLM artifact's lifecycle, capabilities, or context}. This atomic definition guided consistent extraction and enabled subsequent quantitative scoring.

The derivation drew from three complementary streams. \textbf{First}, we systematically reviewed eight influential artifacts in LLM documentation (e.g., Model Cards\cite{mitchell2019model}; see Table\ref{tab:lit-tools-support}). \textbf{Second}, we grounded the inventory in current practice by analyzing the 240-project corpus with a dual strategy: a \textbf{top-down} lens using the 14 principles in Table\ref{tab:guiding_principles}, and \textbf{bottom-up} open coding to surface emergent, practice-based parameters. \textbf{Third}, we normalized the consolidated candidate list to ensure consistency and semantic atomicity:

\begin{itemize}
    \item \textbf{Merging synonyms:} consolidate fields that denote the same concept (e.g., \code{model\_title} and \code{display\_name} $\rightarrow$ \code{model\_name}).
    \item \textbf{Splitting compounds:} decompose broad items into atomic ones (e.g., \code{citation\_info} $\rightarrow$ \code{citation\_authors}, \code{citation\_title}, \code{citation\_year}).
    \item \textbf{Filtering irrelevancies:} remove repository-only metadata unrelated to model documentation (e.g., \code{star\_count}, \code{ci\_badge}).
\end{itemize}

\paragraph{Annotation protocol and quality control}
We used a single-coder protocol with a fixed codebook that defined parameter boundaries and normalization rules. After a cooldown period of at least 14 days, we conducted a stratified intra-rater spot check: a random \textbf{5} sample projects stratified by module and domain, together with 10 projects flagged as borderline in the first pass, were re-examined to verify self-consistency. Any discrepancies triggered codebook refinements and back-propagated edits to earlier items to ensure uniform standards. 

This process yielded a normalized inventory of \textbf{217 atomic parameters}, which forms the evidential bedrock of the hierarchical framework used in the remainder of this paper. The practice-derived, atomic parameterization directly targets \emph{consistency} (normalized field semantics) and \emph{completeness} (exhaustive inventory) to reduce cross-repository fragmentation addressed in RQ1.

\begin{table}[t]
\setlength{\tabcolsep}{4pt}
\renewcommand{\arraystretch}{1.1}
\scriptsize
\centering
\caption{External artifacts informing our parameter inventory. 
Each source highlighted documentation gaps or design practices that were explicitly encoded into the 217 consolidated parameters.}
\label{tab:lit-tools-support}
\begin{tabular}{p{0.23\linewidth} p{0.37\linewidth} p{0.35\linewidth}}
\toprule
\textbf{Source} & \textbf{Highlighted issue / focus} & \textbf{Incorporation into CRAI-MCF parameters} \\
\midrule
Datasheets for Datasets\cite{gebru2021datasheets} & Lack of transparency on data provenance and collection. & Added parameters for data source, collection method, and consent. \\
Model Cards\cite{mitchell2019model} & Inconsistent reporting of intended use and fairness metrics. & Standardized parameters for intended use, out-of-scope use, fairness metrics, and limitations. \\
System Cards\cite{hurst2024gpt} & Absence of societal risk disclosure for LLMs. & Introduced parameters for ethical risks, potential harms, and mitigation strategies. \\
AI Usage Cards\cite{wahle2023ai} & Limited attention to usage-phase accountability. & Added user group, deployment context, and accountability contact. \\
Google Responsible AI Toolkit\cite{google_responsible_ai_toolkit} & Need for operational checklists and governance practices. & Incorporated parameters for governance process, update frequency, and monitoring hooks. \\
Microsoft Responsible AI Dashboard\cite{microsoft_responsible_ai_dashboard} & Emphasis on evaluation transparency and interpretability. & Added parameters for evaluation protocol, explainability notes, and fairness dashboards. \\
Hugging Face Model Card Template\cite{huggingface_model_card_guidebook} & Template-driven documentation with predefined fields. & Used as baseline for parameter extraction; aligned with LLM ecosystem practice. \\
OpenAI System Cards / API Docs\cite{openai_gpt4o_system_card,openai_usage_policies} & Disclosure of limitations, usage policies, and safety mitigations. & Added parameters for risk management, usage constraints, and update/change logs. \\
\bottomrule
\end{tabular}
\end{table}

\subsection{Hierarchical Design}\label{sec:design}

A flat list of 217 parameters, while comprehensive, is impractical for real-world use due to information overload. To turn this inventory into an operational artifact, we thus develop a hierarchical taxonomy. A challenge is to balance coverage with cognitive ergonomics so that the structure remains logically coherent and easy to navigate in day-to-day engineering workflows. Importantly, this is a rules-based, reproducible assignment process rather than ad hoc judgment. Our annotation process maintains an edge-case log and self-consistency checks. Section \ref{sec:evaluation} reports readability and navigability gains under this structure.

\paragraph{Design objectives and constraints.}
We set three objectives for the taxonomy: \emph{navigability} in routine engineering (minimize context switches to find a parameter), \emph{comparability} across projects (non-overlapping module scopes by purpose), and \emph{accountability} for governance reviews (stable structure under incremental updates). These objectives guided both the choice of top-level modules and the granularity of presentation within each module.

\begin{figure*}[t]
    \centering
    \includegraphics[width=0.9\textwidth]{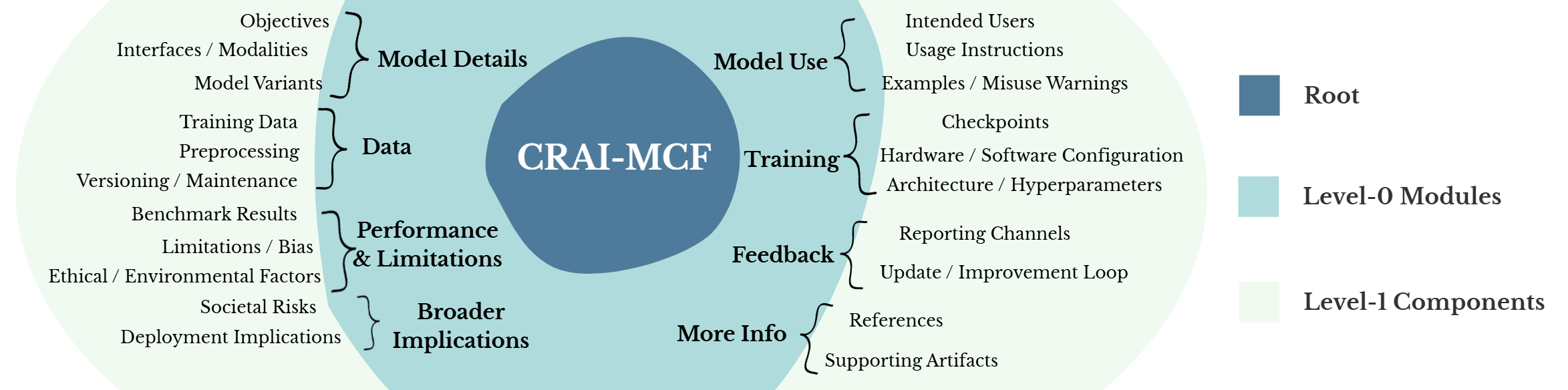}
    \caption{
        Schematic hierarchy of CRAI-MCF. The center node denotes the framework root, the middle band shows the eight fixed Level-0 modules, and the outer band illustrates representative Level-1 components. Deeper levels in the full containment hierarchy are omitted for readability; the figure is illustrative rather than exhaustive.
    }
    \label{fig:hierarchy}
\end{figure*}

\paragraph{Criteria-driven grouping.}
We adopted a \emph{criteria-driven} grouping rather than algorithmic clustering to maximize usability during reviews. Parameters were assigned to modules using three decision rules: 
(1) \emph{function of use} — which downstream decision the parameter supports; 
(2) \emph{lifecycle phase} — data, training, evaluation, deployment/governance;  
(3) \emph{technical vs.\ governance axis} — "technical" covers data/training/performance and reproducibility; "governance" covers intended/misuse, accountability, risk/impact, and feedback mechanisms.  
When rules conflict, we prioritize the parameter’s primary \emph{function of use} to keep modules mutually exclusive by purpose and to minimize cross-referencing cost. The resulting modules are listed in Table\ref{tab:modules}; a schematic tree of the hierarchy appears in Figure\ref{fig:hierarchy}.
\footnote{We have explored parameter clustering approach (e.g., HAC/Leiden on co-occurrence/embeddings), but the resulting clusters tend to merge technical and governance dimensions and were unstable under small corpus changes. We therefore adopted a rules-based design to ensure operational stability.}

To have a pragmatic balance between coverage and cognitive load, finally we chose 8 Level-0 top-level modules. Fewer modules forced heterogeneous concerns into the same bucket (hurting findability), while more modules increased cross-references (hurting flow). Pilot walkthroughs with practitioners indicated that eight preserved coherent scopes along the technical–governance split and remained compact enough for audits. We do not claim an intrinsic optimum; rather, 8 proved stable across domains in our corpus. The eight fixed Level-0 modules and containment-only hierarchy are designed to lower cognitive load (findability, fewer context switches) and to enable \emph{comparability} by keeping module scopes non-overlapping and reviewable side-by-side.

\paragraph{Multi-level structure and information density.}
Only the level-0 top-level \emph{modules} (eight) are fixed and uniquely defined. Below the Level-0 modules, the taxonomy forms a containment hierarchy of \emph{up to five} display levels (Levels 1–5). In practice, many high-salience leaves (by $s_i$) appear near the top of each module page for quick coverage checks, while deeper branches capture the long tail and operational specifics. This variable-depth, containment-only design avoids duplication, keeps modules scannable at a glance, and still supports drill-down to exhaustive detail when needed.

\newcolumntype{L}[1]{>{\raggedright\arraybackslash}m{#1}}

\begin{table}[t]
\setlength{\tabcolsep}{4.5pt}
\renewcommand{\arraystretch}{1.2}
\scriptsize
\centering
\caption{Eight CRAI-MCF modules with aligned value domains and industrial use cases. Representative parameters are shown as supporting evidence rather than exhaustive lists.}
\label{tab:modules}
\begin{tabular}{L{0.22\linewidth} L{0.22\linewidth} L{0.40\linewidth}}
\toprule
\textbf{Module} & \textbf{Aligned value domain (VSD)} & \textbf{Industrial relevance / Example parameters} \\
\midrule
\multirow[c]{2}{*}{Model Details} & Usability    & \multirow[c]{2}{*}{\parbox[t]{\hsize}{Discoverability and licensing clarity (e.g., terms of use, objective function)}} \\
                                  & Transparency & \\
Model Use & Transparency & Scope boundaries for deployment (e.g., intended use, misuse/out-of-scope) \\
\multirow[c]{2}{*}{Data} & Transparency   & \multirow[c]{2}{*}{\parbox[t]{\hsize}{Provenance and consent (e.g., dataset sources, preprocessing)}} \\
                         & Accountability & \\
Training & Sustainability & Reproducibility and efficiency (e.g., hyperparameters, compute resources) \\
Performance \& Limitations & Accountability & Benchmarks with risks (e.g., metrics, fairness/robustness, environmental impact) \\
\multirow[c]{2}{*}{Feedback} & Interactivity & \multirow[c]{2}{*}{\parbox[t]{\hsize}{Continuous improvement (e.g., feedback channels, incident/error reporting)}} \\
                             & Accountability & \\
\multirow[c]{2}{*}{Broader Implications} & Sustainability & \multirow[c]{2}{*}{\parbox[t]{\hsize}{Societal risks and lessons (e.g., ethical concerns, foreseeable impacts)}} \\
                                         & Ethical values & \\
\multirow[c]{2}{*}{More Info} & Usability    & \multirow[c]{2}{*}{\parbox[t]{\hsize}{Extended materials (e.g., configs, seeds, scripts, references)}} \\
                              & Transparency & \\
\bottomrule
\end{tabular}
\end{table}

\subsection{Scoring System}\label{sec:scoring}
Checklist-style documentation provides prompts but does not answer two practical questions for engineering teams: (i) \emph{how much is sufficient}, and (ii) \emph{what to fill first when resources are limited}. CRAI-MCF adopts a concise, reproducible scheme that enables comparability across projects while yielding actionable priorities.

\subsubsection{Parameter-level prior}
For each atomic parameter $p_i$, we assign a reference prior (i.e., a fixed \emph{reference weight} derived from corpus frequency)
\[
s_i=\frac{f_i}{N},
\]
where $f_i$ is the number of projects (out of $N{=}240$) that document $p_i$. We use "prior" solely to mean this reference weight; it does \emph{not} indicate whether a parameter is documented in a given project. The prior captures community salience—frequently reported parameters carry more prior mass—and is \emph{not} a normative importance score. In practice, $s_i$ provides a natural ordering for "what to fill first."

\subsubsection{Module baseline (operational heuristic)}
Let $S_M=\sum_{i\in M}s_i$ be the total attainable prior for module $M$. We define a baseline threshold
\[
\mathrm{BaselineScore}(M)=\Big(\frac{O_M}{O_{\text{All}}}+\frac{A_M}{A_{\text{All}}}\Big)\cdot \frac{S_M}{2},
\]
where $O_M/O_{\text{All}}$ reflects \emph{observed coverage in practice} (how often parameters in $M$ are documented across the corpus), and $A_M/A_{\text{All}}$ reflects the module's \emph{design capacity} (its share of the overall parameter space). The constant $1/2$ places the threshold in a stable, interpretable band within $(0,S_M]$.

Within a project, treat each parameter as documented or not documented. A parameter is counted as documented if it contains substantive content or a verifiable evidence link (placeholders do not count). Compute the \emph{cumulative prior of documented parameters} by summing $s_i$ over the parameters in $M$ that are present \emph{(i.e., add up the reference weights $s_i$ only for the parameters that are marked as present)}. A module $M$ is deemed \emph{sufficient} when this cumulative prior meets or exceeds $\mathrm{BaselineScore}(M)$. Intuitively, a module passes when the documented parameters collectively cover enough of the prior mass to be credible relative to both community practice and the module's design space.

\paragraph{Reducing authoring burden.}
Prior scores allow teams to cover the highest-yield parameters first. Filling fields in descending $s_i$ reaches module sufficiency with fewer entries; lower-$s_i$ items can be deferred to progressive enrichment, cutting time-to-sufficiency and easing maintenance (\textbf{RQ3}).

\paragraph{Comparability under a unified scheme.}
Because priors $s_i$ and module baselines are fixed across projects, CRAI{-}MCF provides a single, role-agnostic score per module, enabling like-for-like cross-model comparison without bespoke templates (\textbf{RQ2}).

\paragraph{Illustrative example.}
Suppose $S_M{=}2.4$, with $O_M/O_{\text{All}}{=}0.18$ and $A_M/A_{\text{All}}{=}0.14$. The baseline is
$\mathrm{BaselineScore}(M)=(0.18+0.14)\cdot(2.4/2)=0.384$.
Assume the project has documented three \textbf{parameters} whose priors are $\{0.12,0.09,0.05\}$, so the cumulative prior is $0.26$ and the shortfall is $0.124$. Filling the next two highest-prior \textbf{parameters} ($0.08$ and $0.06$) raises the cumulative prior to $0.40$, which exceeds the baseline ($0.40\ge0.384$), hence the module becomes sufficient.

\section{Experiment Setup}
\label{sec:setup}
We evaluate CRAI-MCF in realistic software engineering settings using a corpus-based diagnostic, a pilot practitioner study, and a cost assessment. The study is designed to answer the three research questions (RQ1--RQ3) stated in the introduction. A traceability map from questions to metrics and components appears in Table\ref{tab:rq-trace}.

\subsection{Evaluation Metrics and Traceability}
We operationalize the research questions with five metrics: \emph{technical soundness}, \emph{readability}, \emph{completeness}, \emph{applicability}, and \emph{dynamic maintenance}. These metrics align with ISO/IEC~25010 (software quality characteristics) \cite{estdale2018applying}, ISO/IEC~25002:2024 (quality model construction) \cite{iso25002_2024}, the OECD AI Principles, and the NIST AI Risk Management Framework \cite{ai2023artificial}.\par
\indent \textit{\textbf{RQ1}} is assessed through \emph{\textbf{completeness}} and indicators of stable behavior under \emph{\textbf{technical soundness}}.\par
\indent \textit{\textbf{RQ2}} is assessed through \emph{\textbf{readability}} and cross-domain \emph{\textbf{applicability}}.\par
\indent \textit{\textbf{RQ3}} is assessed through \emph{\textbf{dynamic maintenance}}, using an authoring-effort proxy that remains within this metric.\par

\begin{figure}[t]
    \centering
    \includegraphics[width=0.7\linewidth]{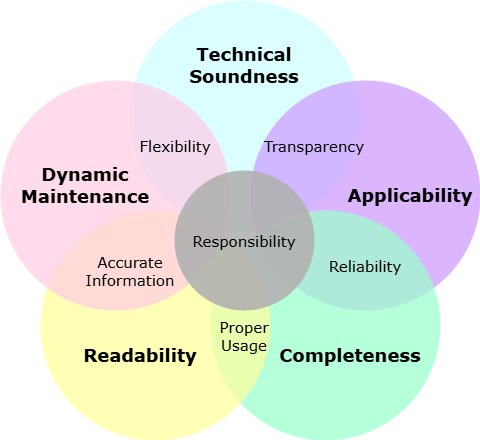}
    \caption{Conceptual integration of the five evaluation dimensions. Technical soundness, readability, completeness, applicability, and dynamic maintenance converge on \textit{responsible} documentation.}
    \label{fig:rubric}
\end{figure}

Table\ref{tab:rubric-mapping} summarizes the standards linkage for the five metrics. Figure\ref{fig:rubric} gives a visual overview. Table\ref{tab:rq-trace} lists the one-to-one mapping from research questions to metrics, components, instruments, and outputs.

\begin{table}[t]
\centering
\caption{Mapping of the five metrics to established standards and governance frameworks.}
\label{tab:rubric-mapping}
\scriptsize
\begin{tabular}{p{2.6cm} p{5.6cm}}
\toprule
\textbf{Metric} & \textbf{Aligned standards/framework concepts} \\
\midrule
Technical soundness & ISO/IEC~25010 (Reliability; Functional suitability); NIST AI RMF (Robustness) \\
Readability & ISO/IEC~25010 (Usability); ISO/IEC~25002:2024 (documentation clarity); OECD (Transparency) \\
Completeness & ISO/IEC~25002:2024 (coverage and sufficiency); NIST AI RMF (documentation sufficiency) \\
Applicability & NIST AI RMF (contextualization of use); OECD (human oversight; robustness in context) \\
Dynamic maintenance & ISO/IEC~25002:2024 (maintainability across lifecycle); NIST AI RMF (continuous monitoring) \\
\bottomrule
\end{tabular}
\end{table}

\begin{table}[t]
\centering
\caption{Traceability from research questions to metrics, components, instruments, and primary outputs.}
\label{tab:rq-trace}
\scriptsize
\begin{tabular}{p{0.3cm} p{1.2cm} p{1.3cm} p{1.6cm} p{2.7cm}}
\toprule
\textbf{RQ} & \textbf{Metrics} & \textbf{Component} & \textbf{Instruments / Tasks} & \textbf{Primary outputs} \\
\midrule
RQ1 & completeness; technical soundness & Coverage Diagnostic & 80-project mapping to modules & coverage profiles; indicators of reduced fragmentation; governance gaps \\
RQ2 & readability; applicability & Pilot Practitioner Study & Likert items; three A/B case tasks & preference for comprehension and consultation; perceived load; exact binomial tests \\
RQ3 & dynamic maintenance & Cost Assessment & information-parity conversion; effort self-report & average prose reduction; update localization; perceived cost delta \\
\bottomrule
\end{tabular}
\end{table}

\subsection{Evaluation Components}
We use three complementary components. Each component provides evidence for one or more research questions.

\paragraph{Coverage Diagnostic (addresses RQ1; informs RQ2).}
We curate an independent validation corpus of 80 high-adoption LLM projects across code generation, healthcare, finance, and education. For each project, we map documentation to CRAI-MCF modules and produce quantitative coverage profiles. The diagnostic measures coverage, surfaces fragmentation, and checks the stability of module patterns across domains.

\paragraph{Pilot Practitioner Study (addresses RQ2).}
We conducted a pilot practitioner study using a structured instrument with three parts. The first part collects Likert ratings on hierarchy clarity, the scoring workflow, and perceived reading speed. The second part asks participants to rank modules under scenarios such as compliance review, research reuse, and customer delivery. The third part collects open-ended feedback on strengths, weaknesses, and improvements. The instrument is designed to be completed within fifteen minutes. The goal was to obtain early practitioner-facing evidence on readability, comparability, and perceived maintenance practicality, rather than to estimate population prevalence.

\paragraph{Cost Assessment (addresses RQ3).}
We estimate effort within the dynamic maintenance metric. Creation effort is proxied by prose length under information parity. Maintenance effort is assessed by the modular structure of CRAI-MCF relative to ad hoc practice. Participants also report whether CRAI-MCF is more, less, or equally costly than current methods.

\subsection{Participants and Procedure}
We recruit 23 volunteers from industry and academia across the United States, Australia, and China. Industrial participants are developers who use or integrate LLM systems in practice. Academic participants have experience in LLM development and evaluation.

Each participant follows a three-step protocol. They review a short briefing with a documentation overview and a demo webpage. They complete the study instrument and provide comparative feedback on cost and maintainability. We anonymize responses. Quantitative data are analyzed with descriptive statistics and hypothesis testing. Qualitative responses are thematically coded to surface recurrent themes.

Although the sample size ($n{=}23$) is modest, recruitment targeted participants with direct experience in LLM use or development and excluded peripheral roles. The practitioner component was designed as a pilot study to gather early evidence on usability, comparability, and perceived maintenance practicality rather than to estimate population prevalence. Accordingly, findings from this component should be interpreted as preliminary practitioner-facing evidence. Subgroup comparisons are exploratory due to limited per-group sizes.

\section{Evaluation} \label{sec:evaluation}
Building on the experimental setup in Section\ref{sec:setup}, this section reports evidence for our three research questions (RQ1-RQ3). Findings are organized by the three study components introduced earlier: a coverage diagnostic on an independent corpus (answers RQ1; informs RQ2), a pilot practitioner study (evidence for RQ2), and a cost assessment (answers RQ3).

\subsection{Coverage Diagnostic}
\label{sec:coverage-diagnostic}
\begin{figure}[t]
  \centering
  \includegraphics[width=\linewidth]{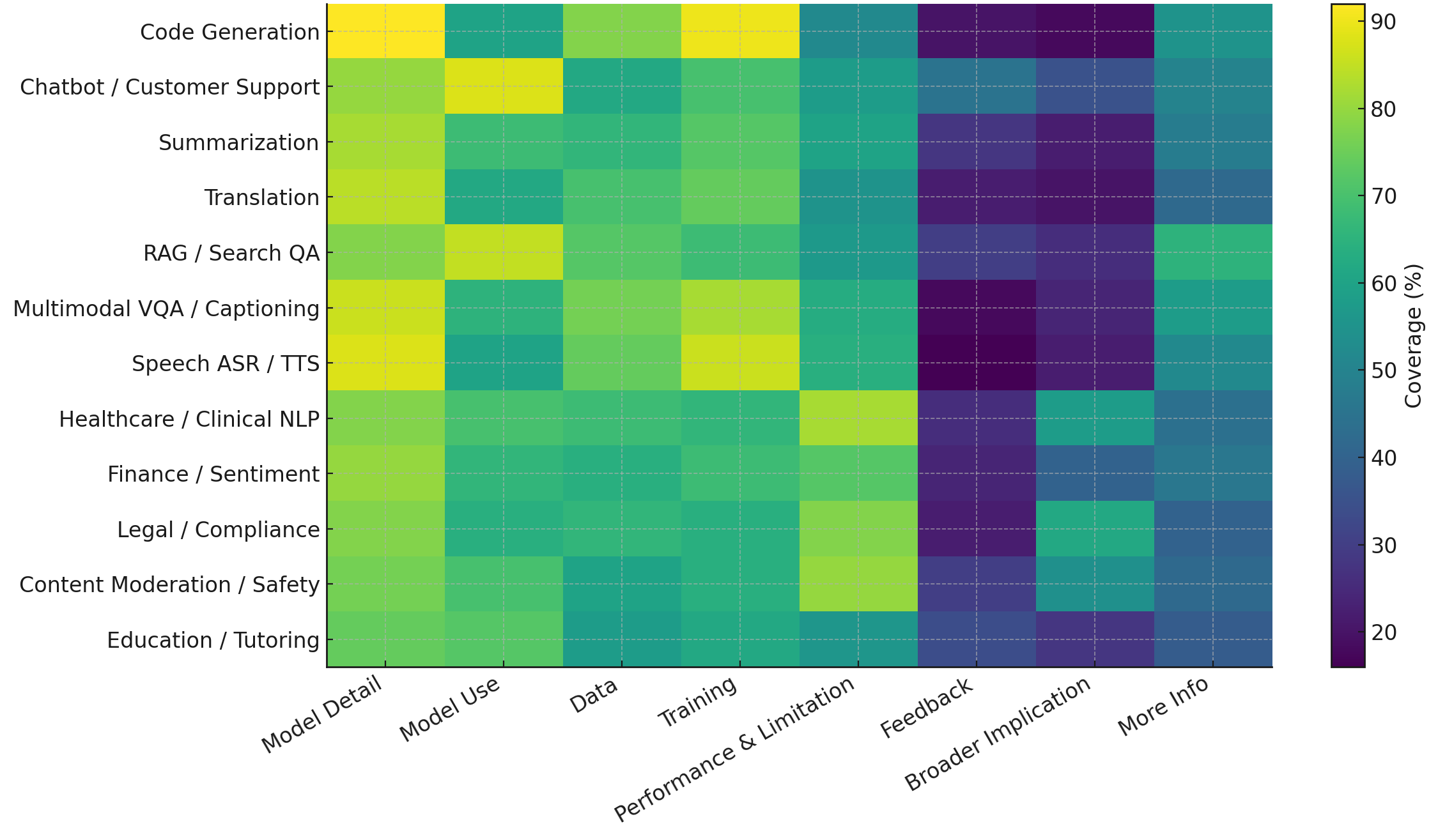}
  \caption{Task-module documentation heatmap. Lighter/warmer cells denote higher coverage; darker/cooler cells denote lower coverage. Persistent dark bands on \emph{Feedback} and, for many tasks, \emph{Broader Implications} indicate ecosystem-level under-reporting of accountability and societal-risk information.}
  \label{fig:heatmap}
\end{figure}

Figure\ref{fig:heatmap} maps task families (rows) to CRAI-MCF’s eight modules (columns). The colorbar denotes module-level parameter coverage (0-100\%), computed as the fraction of parameters present for a task-module pairing relative to that module’s parameter set,
\[
d_{t,M}=\frac{|P_{t}\cap P_{M}|}{|P_{M}|},
\]
where $P_{t}$ denotes the set of CRAI-MCF parameters documented for task family $t$, and $P_{M}$ denotes the parameter set defined by module $M$.

\subsubsection{Salient patterns.}
Three structural regularities are consistent across tasks:

\paragraph{Technical vs.\ governance.}
Pre-deployment technical descriptors are well covered (\emph{Model Details} high; \emph{Data} and \emph{Training} mid to high). Governance-oriented modules remain low (\emph{Feedback}, \emph{Broader Implications}). Coverage in technical modules does \emph{not} co-vary with governance coverage, indicating separate practices that require distinct prompts and ownership.

\paragraph{Usage and evaluation.}
\emph{Model Use} and \emph{Performance \& Limitations} sit mid-range: intended use and aggregate benchmarks are common, while decision-critical items—misuse/out-of-scope, operational preconditions, fairness/robustness, and uncertainty—are sporadic.

\paragraph{Task-cluster effects without parity.}
Regulated domains (\emph{Healthcare / Clinical NLP}, \emph{Legal / Compliance}) raise governance and evaluation columns relative to general tasks, and interaction-heavy workloads (\emph{Chatbot / Customer Support}, \emph{RAG / Search QA}) raise \emph{Model Use}. In all cases, governance still trails the same row’s technical columns; none raises coverage in \emph{Feedback} materially.

\subsubsection{Answer for RQ1.}
Assessed via the \emph{completeness} and \emph{technical soundness} metrics, we find:

\textit{Completeness.} The heatmap converts narrative omissions into measurable, module-level gaps. Governance shortfalls are localized to concrete parameters (e.g., incident reporting, uncertainty notes), while technical modules are comparatively well covered. This makes missing content auditable rather than diffuse.

\textit{Technical soundness.} The relative ordering of module coverage is stable across task families and domains in Figure\ref{fig:heatmap}: technical modules remain high, usage and evaluation remain mid-range, and governance modules remain low. This stable behaviour indicates that the practice-derived taxonomy yields repeatable coverage profiles across heterogeneous tasks, improving technical soundness while reducing fragmentation.

\begin{tcolorbox}[colback=white,colframe=black!60,boxrule=0.5pt,arc=2pt,left=6pt,right=6pt,top=4pt,bottom=4pt]
\textbf{Answer for RQ1}: A practice-driven taxonomy resolves documentation fragmentation by converting omission into measurable, auditable gaps and stabilizing coverage profile across domains.
\end{tcolorbox}

\subsection{Pilot Practitioner Study}
\label{sec:survey}
We conducted a pilot practitioner study to gather preliminary evidence on the \emph{readability} and \emph{applicability} of CRAI-MCF.
The study used a structured instrument with 3 components:
(i) three A/B case tasks where participants compared the original documentation with a CRAI-MCF version for \emph{general}, \emph{medical}, and \emph{creative} LLMs (stimuli in Table\ref{tab:stimuli}); 
(ii) judgments on procedural clarity and cross-module comparability; and 
(iii) a choice of preferred parameter granularity.

\subsubsection{Salient patterns.}
\begin{figure}[t]
  \centering
  \includegraphics[width=0.9\linewidth]{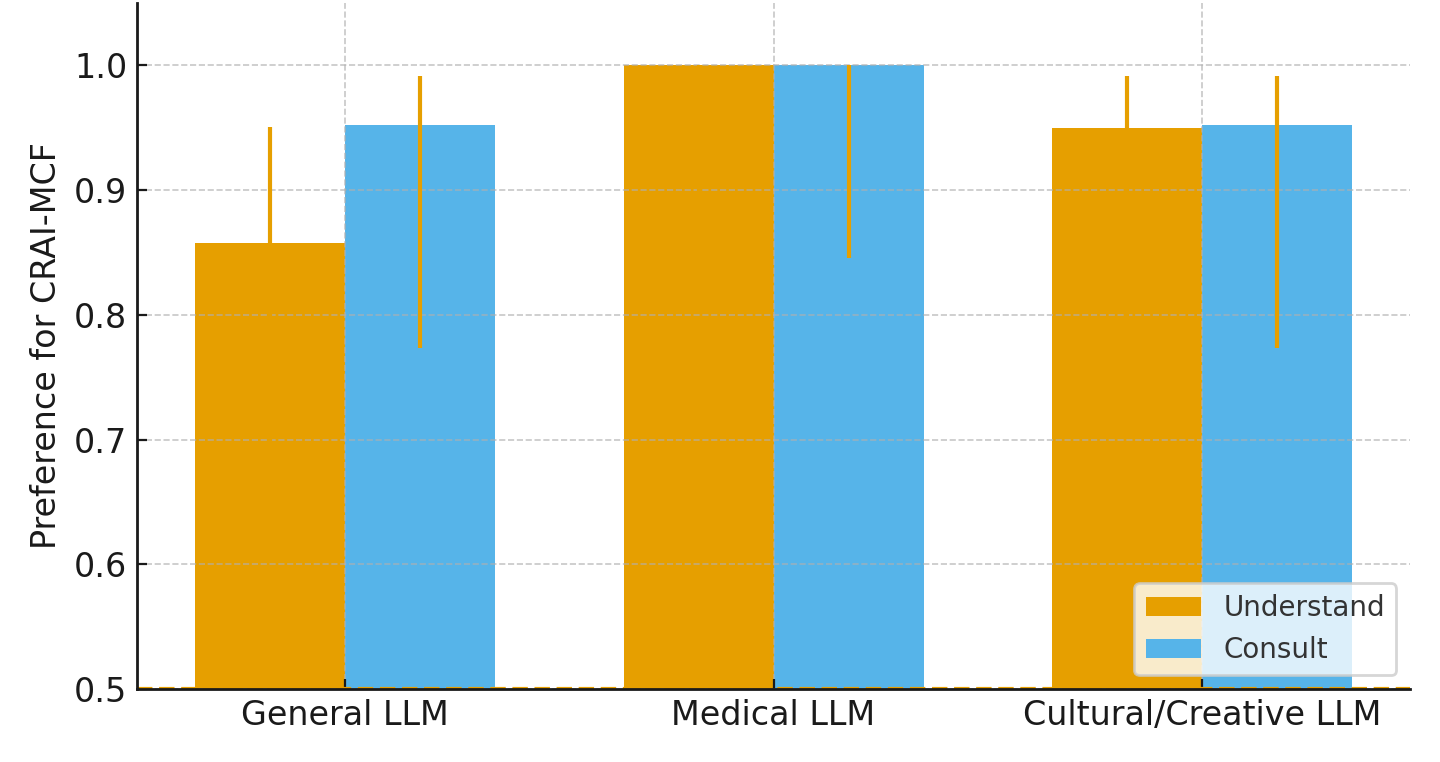}
  \caption{Three-domain case comparison. Bars report the proportion preferring CRAI-MCF over the original for \emph{rapid task comprehension} and \emph{preferred version to consult} (Wilson 95\% CIs; dashed line at 0.5 marks the chance baseline).}
  \label{fig:case-comparison}
\end{figure}

Three consistent patterns emerge across the pilot cases:
\paragraph{Structure-driven usability gains.} Across all 3 domains, participants in the pilot study favoured the CRAI-MCF variant for both \emph{rapid task comprehension} and \emph{which version they would consult}. Table\ref{tab:ab-forest} reports the A/B outcomes with Wilson 95\% confidence intervals: general (18/21; 20/21), medical (21/21; 21/21), and creative (19/20; 20/21). Pooled preference is 93.5\% (95\% CI [84.6, 97.5]) for comprehension and 96.8\% ([89.1, 99.1]) for consultation. In parallel workflow-level judgments, respondents unanimously endorsed both statements—"the scoring workflow improves procedural clarity" and "multi-level scoring improves cross-module comparability" (23/23; five-point Likert collapsed to Agree/Neutral/Disagree).
\paragraph{Cross-domain implementability without bespoke templates.} The same hierarchy yields consistent wins under heterogeneous constraints—utility oriented (general), compliance-heavy (medical), and content-centric(creative), supporting transferability by \emph{reweighting what to read next} instead of redesigning templates per domain. Open-ended remarks frequently mentioned "clear structure", "systematic organization", and "reduced reading pressure", in line with the quantitative A/B outcomes (Fig.\ref{fig:case-comparison}).

\paragraph{Evidence-oriented completeness with targeted gaps.} Practitioners generally prefer higher detail (Table\ref{tab:granularity}: L4--L5 $=15/23=65.2\%$, Wilson 95\% CI $[44.9, 81.2\%]$), and none select "not sufficient". Rather than requesting entirely new dimensions, respondents mainly asked for more explicit \emph{evidence slots} in low-coverage areas—such as security/privacy drills, uncertainty indicators, and machine-linkable artifacts—mirroring corpus-level troughs in \emph{Feedback} and parts of \emph{Performance \& Limitations}.

\begin{table}[t]
\centering
\begin{threeparttable}
\caption{Stimuli used in the A/B case tasks (exact identifiers for replication).}
\label{tab:stimuli}
\scriptsize
\begin{tabular}{@{}L{2cm} l l l@{}}
\toprule
Case & Domain & Artifact (snapshot) & Repo Link \\
\midrule
Dolphin Mistral\tnote{a} & General LLM & Model card (2025-09-08) &
\href{https://huggingface.co/dphn/Dolphin-Mistral-24B-Venice-Edition}{link} \\
BioGPT & Medical LLM & Model card (2022-02-03) &
\href{https://huggingface.co/microsoft/biogpt}{link} \\
MI \& Opt & Creative LLM & README (2023-03-24) &
\href{https://github.com/Stry233/MlAndOpt-LLM-for-Creative-Story-Generation}{link} \\
\bottomrule
\end{tabular}
\begin{tablenotes}\footnotesize
\item[a] Full name: \emph{Dolphin Mistral 24B Venice Edition}.
\end{tablenotes}
\end{threeparttable}
\end{table}

\begin{table}[t]
\centering
\caption{A/B preference for CRAI-MCF with Wilson 95\% confidence intervals.}
\label{tab:ab-forest}
\scriptsize
\begin{tabular}{lrrr}
\toprule
Outcome & Success/Total & Proportion & 95\% CI \\
\midrule
Understand: Task A (general) & 18/21 & 85.7\% & [65.4\%, 95.0\%] \\
Understand: Task B (medical) & 21/21 & 100.0\% & [84.5\%, 100.0\%] \\
Understand: Task C (creative) & 19/20 & 95.0\% & [76.4\%, 99.1\%] \\
Consult: Task A (general)    & 20/21 & 95.2\% & [77.3\%, 99.2\%] \\
Consult: Task B (medical)    & 21/21 & 100.0\% & [84.5\%, 100.0\%] \\
Consult: Task C (creative)   & 20/21 & 95.2\% & [77.3\%, 99.2\%] \\
\midrule
Understand: pooled & 58/62 & 93.5\% & [84.6\%, 97.5\%] \\
Consult: pooled    & 61/63 & 96.8\% & [89.1\%, 99.1\%] \\
\bottomrule
\end{tabular}
\end{table}

\noindent\emph{Statistical note.}
One-sided exact binomial tests vs.\ $0.5$ (per task): 
Understand A $p=7.45\times10^{-4}$, B $p=4.77\times10^{-7}$, C $p=2.00\times10^{-5}$; 
Consult A $p=1.05\times10^{-5}$, B $p=4.77\times10^{-7}$, C $p=1.05\times10^{-5}$. 
All $p<.001$. \emph{Method notes:} case/task order and A/B assignment were fixed (no randomization). Responses were anonymous and untimed. These statistics are reported as descriptive support for the pilot study and should not be interpreted as definitive evidence of downstream workflow improvement.

\begin{table}[t]
\centering
\caption{Preferred parameter granularity (n=23).}
\label{tab:granularity}
\scriptsize
\begin{tabular}{lrrrrrr}
\toprule
Level & L1 & L2 & L3 & L4 & L5 & Not sufficient \\
\midrule
Count   & 1 & 2 & 5 & 9 & 6 & 0 \\
Percent & 4.3\% & 8.7\% & 21.7\% & \textbf{39.1\%} & \textbf{26.1\%} & 0.0\% \\
\bottomrule
\end{tabular}
\end{table}

\subsubsection{Answer for RQ2}
The pilot practitioner study provides initial evidence that CRAI-MCF improves \textbf{readability} and perceived \textbf{comparability}. Participants generally preferred the CRAI-MCF structure for rapid comprehension, consultation, and procedural clarity. Pooled A/B preference was 93.5\% for comprehension and 96.8\% for consultation (Table\ref{tab:ab-forest}).

For \textbf{comparability}, all respondents agreed that the multi-level structure improves cross-module comparability (23/23). The fixed eight-module grid and the scoring workflow give a common basis for like-for-like review, so models can be compared at the same module level rather than through heterogeneous narratives. The effect holds across three distinct domains (general, medical, creative).

Participants asked for evidence fields within existing modules (e.g., links to evaluation scripts, configuration hashes, incident reporting, uncertainty indicators) rather than new dimensions. This suggests that finer granularity is acceptable when tied to verifiable artifacts, and points to a practical path for improving completeness by populating evidence in known gaps rather than expanding the schema.

\begin{tcolorbox}[colback=white,colframe=black!60,boxrule=0.5pt,arc=2pt,left=6pt,right=6pt,top=4pt,bottom=4pt]
\textbf{Answer for RQ2}: The pilot practitioner study provides initial evidence that a value-aligned modular architecture with quantitative scoring can reduce perceived cognitive load and support more structured like-for-like comparison across diverse LLMs.
\end{tcolorbox}

\subsection{Cost Assessment}
\label{sec:cost}
We assess whether CRAI-MCF keeps documentation effort within practical limits under the \emph{dynamic maintenance} metric. We use two lenses: (i) an \emph{information-parity} restructuring as a coarse, auditable proxy for reading and coordination cost, and (ii) practitioner perceptions from the pilot study (Section \ref{sec:survey}). The focus is on how much prose teams must produce and consume and on how localized updates become, rather than on a stopwatch estimate.

To mitigate the potential bias, we applied two pipelines to each artifact while preserving all information: an LLM-assisted conversion (GPT-4o with a constrained prompt) and a human conversion following the same checklist. Discrepancies were adjudicated by retaining the union of facts. Counts include prose only (excluding code, tables, and figure captions). Materials comprise the three domain cases (general, medical, creative) and four widely used public artifacts (BLOOM, Transformer, RoBERTa-base, Audio Spectrogram Transformer (AST)).

Across these artifacts, the information-parity conversion yields an average reduction of about 38\% in textual surface for the same information. This represents a substantial decrease in reading and coordination burden and matches the pilot study outcomes in Section \ref{sec:survey} (lower perceived reading pressure, clearer “what to read next,” and preference for rapid comprehension and consultation). Information parity means that every original fact, critical link, and primary evidence statement is preserved and re-mapped into the CRAI-MCF structure by placing content in its corresponding \emph{semantically atomic parameter slot}. This eliminates narrative redundancy and organizational prose.

The observed reductions arise from three main sources: replacing narrative scaffolding with module labels and evidence fields; deduplicating caveats and usage notes into \emph{Model Use} and \emph{Performance \& Limitations}; and externalizing fine-grained detail to machine-linkable evidence pointers.

\paragraph{Limits.}
Word count is a coarse proxy for effort rather than timing estimate. Some reduction is inherent in aligning narrative content to discrete parameter slots. The bias is bounded by the dual-pipeline protocol and explicit information-parity instructions. Further limitations are discussed in Section \ref{sec:threats}. Even so, the $\sim$38\% average reduction is directionally large and consistent across materials.

\subsubsection{Answer for RQ3}
CRAI-MCF can be operated within industrial constraints. First, the information parity conversion shows an average $\sim$38\% reduction in textual surface while preserving every fact, link, and primary evidence item, which keeps documentation quality intact and makes the artifact more systematic by mapping content to semantically atomic slots. Second, maintenance is localized: routine changes concentrate in \emph{Model Details} and \emph{Performance \& Limitations}, where teams update pointers and evidence instead of rewriting narrative. Third, practitioners report lower reading load and clearer update targets, which reduces use and authoring effort. Taken together, these results indicate that authoring and maintenance remain within practical limits while quality is preserved through verifiable evidence attachment, thereby answering RQ3.

\begin{tcolorbox}[colback=white,colframe=black!60,boxrule=0.5pt,arc=2pt,left=6pt,right=6pt,top=4pt,bottom=4pt]
\textbf{Answer for RQ3}: CRAI-MCF can be operationalized within industrial constraints, reducing authoring and maintenance overhead while preserving documentation quality through structured evidence attachment.
\end{tcolorbox}

\section{Threats to Validity}
\label{sec:threats}

\textbf{Internal validity.}
Our evaluation focuses on documentation use rather than live operational workflows. The A/B judgments were based on documentation snapshots rather than documentation embedded in real release or governance systems, so the observed effects may differ under production constraints. In addition, case order and A/B assignment were fixed (no randomization or counterbalancing), so order and learning effects cannot be fully ruled out. We partially mitigated these risks by keeping instructions symmetric, administering responses anonymously with no time budget (Section\ref{sec:survey}), and standardizing stimuli across three domains. Mapping corpus artifacts to CRAI-MCF modules for the coverage diagnostic (Fig.\ref{fig:heatmap}) also involves coder judgment; we used a written mapping guide and spot checks to reduce drift. For the cost analysis (Section\ref{sec:cost}), some residual summarization bias may remain despite adjudication between LLM-assisted and human conversions.

\textbf{Construct validity.}
Our endpoints capture the utility of documentation structure rather than direct improvements in model quality or deployment outcomes. Rapid comprehension and preferred version to consult operationalize readability, but do not directly measure downstream success such as better model selection or governance decisions. Likewise, the coverage metric (Section\ref{sec:coverage-diagnostic}) captures parameter presence rather than evidence quality. The cost proxy is text length under information parity rather than wall-clock effort, and should therefore be interpreted as an operational signal rather than a precise time estimate.

\textbf{External validity.}
The pilot practitioner study sample ($n=23$) is purposive, involving practitioners and researchers with LLM experience across the US, AU, and CN. This supports utility claims for informed users, but limits broader generalization and may underrepresent non-English or organization-specific documentation practices. More importantly, we do not yet report a longitudinal deployment study in a live industrial setting. CRAI-MCF has not been evaluated through sustained use in an enterprise governance process, procurement workflow, or release pipeline. The three A/B cases broaden domain coverage, but cannot span the full design space of LLM deployment, especially in highly regulated or safety-critical settings.

\textbf{Conclusion validity.}
Preference effects were evaluated with Wilson confidence intervals and exact binomial tests; however, multiple comparisons and small per-case counts warrant conservative interpretation. For the cost proxy, reporting a single aggregate reduction (about 38\%) may mask variance across artifacts and workflows. Overall, the current evidence is best interpreted as supporting CRAI-MCF's usefulness for improving documentation navigability, comparability, and maintenance structure, rather than as definitive proof of downstream gains in live industrial decision-making.

\textbf{Ethics statement.}
The pilot practitioner study involved human participants and complied with the ACM Publications Policy on research involving human participants and subjects. Our institution’s ethics procedures determined the study to be minimal risk and exempt from full review. All participants provided informed consent, participation was voluntary and uncompensated, and responses were de-identified prior to analysis. No proprietary or company-confidential artifacts were used.

\section{Conclusion and Future Work}
\label{sec:conclusion}

\paragraph{\textbf{Future work.}}
An important next step is to move from passive documentation assessment to more proactive workflow support. One direction is to assist authors by structuring existing project documentation into CRAI-MCF-aligned templates while preserving verifiable evidence links. Another is to make the framework more adaptive over time: as large volumes of README files, model cards, and related documentation continue to accumulate, documentation trends may reveal newly salient parameters and shifting reporting priorities across the ecosystem. Incorporating such signals could help future versions of CRAI-MCF evolve from a fixed template into a continuously updated documentation framework that remains aligned with emerging industrial practice. A further step is to integrate sufficiency checks into CI/CD and software supply-chain artifacts (e.g., SBOMs and registries), so that documentation quality becomes a built-in property of software delivery rather than an afterthought.

\paragraph{\textbf{Conclusion.}}
CRAI-MCF reframes narrative model cards as an auditable, eight-module hierarchy of 217 atomic parameters with priors and baselines. Evidence from a coverage diagnostic and a pilot practitioner study suggests clearer technical-governance ownership, more structured like-for-like review, and a mean $\sim$38\% prose reduction under information parity without loss of facts. In the pilot study, practitioners strongly preferred CRAI-MCF for rapid comprehension and consultation, providing initial evidence of lower perceived cognitive load and smoother review flows. In practice, adoption is lightweight: \emph{map} existing materials to the grid, reach \emph{module baselines} using high-prior fields, attach \emph{verifiable evidence}, and \emph{gate} releases when sufficiency falls.

\section*{Data Availability}
We release a replication package online\footnote{\url{https://doi.org/10.6084/m9.figshare.31920750}} that contains:

\begin{itemize}
    \item A parameter template file with occurrence frequencies aggregated from 240 projects;
    \item A concise overview slide deck; 
\end{itemize}

In addition, we provide a lightweight demo showcasing our envisioned system \footnote{\url{https://yyue-eric.github.io/model_card_web/}}.

\balance
\bibliographystyle{ACM-Reference-Format}
\bibliography{references}

\end{document}